# Identifying the 'Fingerprint' of Antiferromagnetic Spin-fluctuations on Iron-Pnictide Superconductivity

M.P. Allan[1,2] [§], Kyungmin Lee[1§], A. W. Rost[1,3,4§], M.H. Fischer[1], F. Massee[1,2], K. Kihou[5,6], C.-H. Lee[5,6], A. Iyo[5,6], H. Eisaki[5,6], T.-M. Chuang[7], J.C. Davis[1,2,3,8], and Eun-Ah Kim[1]

1. *LASSP, Department of Physics, Cornell University, Ithaca, NY 14853, USA.*
2. *CMPMS Department, Brookhaven National Laboratory, Upton, NY 11973, USA.*
3. *School of Physics and Astronomy, University of St Andrews, St Andrews, Fife KY16 9SS, Scotland.*
4. *Department of Physics, The University of Tokyo, Bunkyo-ku, Tokyo 113-0033, Japan.*
5. *Institute of Advanced Industrial Science and Technology, Tsukuba, Ibaraki 305-8568, Japan.*
6. *JST, Transformative Research-Project on Iron Pnictides (TRIP), Tokyo 102-0075, Japan.*
7. *Institute of Physics, Academia Sinica, Nankang, Taipei 11529, Taiwan.*
8. *Kavli Institute at Cornell for Nanoscale Science, Cornell University, Ithaca, NY 14853, USA.*

[§] These authors contributed equally to this work

**Cooper pairing in the iron-based high-$T_c$ superconductors[1-3] is often conjectured to involve bosonic fluctuations. Among the candidates are antiferromagnetic spin fluctuations[1-5] and *d*-orbital fluctuations amplified by phonons[6,7]. Any such electron-boson interaction should alter the electron's 'self-energy', and then become detectable through consequent modifications in the energy dependence of the electron's momentum and lifetime [8-10]. Here we introduce a novel theoretical/experimental approach aimed at uniquely identifying the relevant fluctuations of iron-based superconductors by measuring effects of their self-energy. We use innovative quasiparticle interference (QPI) imaging[11] techniques in LiFeAs to reveal strongly momentum-space anisotropic self-energy signatures that are focused along the Fe-Fe (interband scattering) direction, where the spin fluctuations of LiFeAs are concentrated. These effects coincide in energy with perturbations to the density-of-states $N(\omega)$ usually associated with the Cooper pairing interaction. We show that all the measured phenomena comprise the predicted QPI "fingerprint" of a self-energy due to antiferromagnetic spin-fluctuations, thereby distinguishing them as the predominant electron-boson interaction.**

**1**     The microscopic mechanism for Cooper pairing in iron-based high-temperature superconductors has not been identified definitively[1-3]. Among the complicating features in these superconductors is the multiband electronic structure (see Fig. 1a). However, it is believed widely that the proximity to spin order[1-5] and/or orbital order[6,7] plays a key role in the Cooper pairing. In particular, two leading proposals for fluctuation-exchange-pairing mechanisms focus on two distinct bosonic modes associated with specific broken-symmetry states: anti-ferromagnetic spin fluctuations carrying momentum $\mathbf{Q}=(\pi,\pi)/a_0$, and $d$-orbital fluctuations caused by $E_g$-phonon lattice vibrations of the Fe ions. No conclusive evidence that either fluctuation couples strongly to electrons and is thus relevant to Fe-based superconductivity has been achieved within the plethora of proposals about the existing data[12-18].

**2**     Each type of electron-boson interaction should produce a characteristic electronic 'self-energy' $\hat{\Sigma}(\mathbf{k},\omega)$ representing its effect on every non-interacting electronic state $|\mathbf{k}\rangle$ with momentum $\hbar\mathbf{k}$ and energy $\hbar\omega$. Thus, the interacting Green's function $\hat{G}(\mathbf{k},\omega)$ is given by

$$[\hat{G}(\mathbf{k},\omega)]^{-1} = [\hat{G}^0(\mathbf{k},\omega)]^{-1} - \hat{\Sigma}(\mathbf{k},\omega) \quad , \qquad (1)$$

where $\hat{G}^0(\mathbf{k},\omega)$ represents non-interacting electrons and the detailed structure of $\hat{\Sigma}(\mathbf{k},\omega)$ encapsulates the Cooper pairing process. Here, a hat $\hat{\ }$ denotes a matrix in particle-hole space (Nambu space) for Bogoliubov quasi-particles in the superconducting state. The real part $\text{Re}\hat{\Sigma}(\mathbf{k},\omega)$ then describes changes in the electron's dispersion $\mathbf{k}(\omega)$ and the imaginary part $\text{Im}\hat{\Sigma}(\mathbf{k},\omega)$ describes changes in its inverse lifetime $\tau^{-1}(\mathbf{k},\omega)$. The simplest diagrammatic representation of this electron-boson interaction is shown in Fig. 1b. One way to detect the experimental signature of such a self-energy is to use angle-resolved photo-emission spectroscopy (ARPES) to measure the spectral function $A(\mathbf{k},\omega) \propto \text{Im}G(\mathbf{k},\omega)$ of the states with $\omega<0$. However, it has recently been realized that quasiparticle interference imaging, which can access momentum-resolved information of both filled and empty states with excellent energy resolution ($\delta\omega<0.35$meV at $T=1.2$K), might prove especially advantageous for detecting self-energy effects[19]. Our QPI data are obtained by first visualizing scattering interference patterns in real-space (**r**-space) images of the tip-sample differential tunneling conductance d$I$/d$V$(**r**,$\omega$=$eV$)≡$g$(**r**,$\omega$) using spectroscopic-imaging scanning tunneling microscopy, and then Fourier transforming $g$(**r**,$\omega$) to obtain the power spectral density $g$(**q**,$\omega$)[11]. The $g$(**q**,$\omega$) can then be used to reveal the electron dispersion **k**($\omega$) because elastic scattering of electrons from −**k**($\omega$) to +**k**($\omega$) results in high intensity at **q**($\omega$)=2**k**($\omega$) in $g$(**q**,$\omega$). Sudden changes in the energy evolution **k**($\omega$) due to $\Sigma$(**k**,$\omega$) can then be determined, in principle[19], using such data.

**3**     In a conventional single band s-wave superconductor with isotropic energy gap magnitude $\Delta$, it has been well-established that coupling to an optical phonon with frequency $\Omega$ can lead to a renormalization of the electronic spectra at energy $\Delta$+$\Omega$ ($\hbar$=1) due to a singularity in the momentum independent self-energy $\Sigma$(**k**,$\omega$)=$\Sigma$($\omega$) at $\omega$=$\Delta$+$\Omega$[20]. This classic case is illustrated in Fig. 1c,d through a model spectral function $A$(**k**,$\omega$)∝Im$G$(**k**,$\omega$) and the associated density of states $N$($\omega$)=∫ d**k** $A$(**k**,$\omega$). In Fig. 1c, the "free" dispersion of a hole-like band is represented by the red dashed line, while the renormalized dispersion **k**($\omega$) due to $\Sigma$($\omega$) is highlighted by the locus of maxima in $A$(**k**,$\omega$). These effects can be understood from the conservation of energy and momentum during scattering processes (Fig. 1b), where the flat dispersion of an optical phonon presents constraints only on energy without any momentum dependence.

**4**     In developing our new approach to "fingerprinting" different electron-boson interactions using QPI, we use the realization that the kinematic constraints for a multi-band electronic system coupled to resonant AFSF with a sharp momentum structure should result in a strongly momentum-dependent (anisotropic) self-energy. This is because, given a fermionic dispersion $(\mathbf{k}, \omega_{\mathbf{k}}^n)$ for different bands $n$ and a spectrum of spin fluctuations whose intensity is strongly concentrated at (**Q**,$\Omega$), the renormalization due to the self-energy at a point $(\mathbf{k}, \omega_{\mathbf{k}}^n)$ will be most intense when that point can be connected to another point $(\mathbf{k} - \mathbf{Q}, \omega_{\mathbf{k}-\mathbf{Q}}^m)$ on a different band $m$, such that

$$\omega_{\mathbf{k}}^n = \omega_{\mathbf{k}-\mathbf{Q}}^n - \Omega. \qquad (2)$$

This is the constraint from conservation of both energy and momentum in the electron-

AFSF interaction and its consequence is shown schematically in Fig. 1e. Here the blue(yellow) surfaces represent the hole(electron) bands. The transfer of momentum $\mathbf{Q}=(\pi,\pi)/a_0$ and energy $\Omega$ necessary for the resonant antiferromagnetic fluctuation to couple these bands can be analyzed by shifting the electron-pocket-dispersion surface (horizontally) by $\mathbf{Q}$ and (vertically) by $\Omega$ in the $\mathbf{k}$-$\omega$ space, to obtain the transparent red surface. The black curve, showing the intersection of this red surface with central $\gamma$ band dispersion (blue), is where the kinematic constraint of equation (2) can be satisfied and thus where the strongest self-energy effect due to coupling to AFSF is predicted. The resulting strongly anisotropic renormalization due to electron-AFSF coupling is in strong contrast to what is expected as a consequence of the electron-phonon coupling case discussed above.

**5** Here we study the representative iron-based superconductor LiFeAs as a concrete example for which it should be possible to make a clear theoretical distinction between the self-energy effects driven by different types of bosonic fluctuations. We assume that BCS theory adequately describes the superconductor deep in the superconducting phase. Hence, the non-interacting Green's function is given by

$$[\hat{G}^0(\mathbf{k},\omega)]^{-1} = \omega\hat{\tau}^0 - \Delta_\mathbf{k}\hat{\tau}^1 - H^0_\mathbf{k}\hat{\tau}^3 \quad , \qquad (3)$$

where $\hat{\tau}^0$ and $\hat{\tau}^i$ are the identity and the Pauli matrices in Nambu space, respectively. The superconducting gap structure $\Delta_\mathbf{k}$ and the band structure $H^0_\mathbf{k}$ are taken from experiments[11,12,17] and *ab-initio* calculations[21] (Supplemental Information (SI) Section I). We then study the lowest order self-energy due to the coupling between Bogoliubov quasiparticles and two bosonic modes: a resonant AFSF[22,23] and an optical phonon of the type driving orbital fluctuations due to in-plane lattice vibrations of the Fe ions with $E_g$ symmetry (Fe-$E_g$ phonon). It is the coupling of this Fe-$E_g$ phonon to electrons that is proposed to enhance the *d*-orbital fluctuations which mediate Cooper pairing in the orbital fluctuation mechanism[6,7]. We take a perturbative approach of computing the self-energy to the lowest order[9] (SI Section II):

$$\hat{\Sigma}^{(1)}_{mn}(\mathbf{k},\omega) = \int d\mathbf{q}\, d\nu\, D(\mathbf{q},\nu)\hat{g}^{ml}\hat{G}^0_{ll'}(\mathbf{k}-\mathbf{q},\omega-\nu)\hat{g}^{l'n} \quad , \qquad (4)$$

where the repeated indices are summed over. Given independent quantitative knowledge of the gap structure, such a perturbative treatment can accurately capture the salient features of renormalization due to electron-boson coupling (SI Section III). In equation (4), the bosonic Green's function $D(\mathbf{q},v)$ is sharply peaked around $\mathbf{Q}=(\pi,\pi)/a_0$ with the characteristic energy of $\Omega\approx6$meV to model the resonant AFSF of LiFeAs[22,23], while it is nearly momentum independent for the optical Eg phonon[20]. We focus on the self-energy effects on the γ band (Fig. 1a,e) in the rest of this paper, as its nearly uniform orbital character ($d_{xy}$) greatly simplifies the theoretical study (see SI Section III) while at the same time being readily accessible to QPI studies[11]. Given the geometry of the Fermi surfaces, the kinematic constraint for coupling to resonant AFSF with momentum $\mathbf{Q}$ and energy $\Omega$ (red arrows in Fig. 1e) connects a given $(\mathbf{k},\omega_{\mathbf{k}}^{\gamma})$ on the γ band (blue surface in Fig. 1e) to a point with momentum $\mathbf{k}-\mathbf{Q}$ on one of the two electron-like bands (yellow surfaces in Fig.1e). Thus, the distinct anisotropic dispersions of each band mean that resonant AFSF should result in self-energy effects with a strong directional dependence (black curve on γ band in Fig. 1e). Similarly, for the Fe-$E_g$ phonons with a weak momentum dependence[7], the self-energy effect for the γ band (which consists almost entirely of $d_{xy}$ orbitals[24]) is predicted to be angle-independent (SI Section IV).

6    In Fig. 2a-d we present the predictions from equation (4) for $g(\mathbf{q},\omega)$ in LiFeAs, in the presence of self-energy effects due to coupling to AFSF (SI Sections IV,V). Just below the maximum gap value on the γ band of 3meV (Fig. 2a), the high-intensity region around $q \approx 2k_F^{\gamma}$ shows an anisotropy dictated by the gap anisotropy[11,17,25] with the QPI intensity suppressed along the gap maximum (Fe-As) direction. At energies exceeding the maximum gap values, the predicted $g(\mathbf{q},\omega)$ at first becomes isotropic (Fig. 2b) as one might expect from the fact that the Bogoliubon energy is dominated by the kinetic energy over the gap at high energies. However, at energies $\omega\geq12$meV the predicted self-energy effects for the AFSF self-energy (Fig. 2c,d) are seen and, in fact, strongly suppress the $g(\mathbf{q},\omega)$ intensity in the Fe-Fe direction relative to the Fe-As direction. The complete predicted evolution of $g(\mathbf{q},\omega)$, from being dominated by the anisotropic gap structure[11] to the new effects of the AFSF-driven $\Sigma(\mathbf{k},\omega)$ introduced here, is shown in the left panels of the SI movie M1.

**7** The experimental search for such signatures of $\Sigma(\mathbf{k},\omega)$ in QPI data consists of imaging $g(\mathbf{r},\omega)$ at $T$=1.2K with 0.35meV energy resolution on LiFeAs samples exhibiting $T_c$≈15K and with the superconducting energy gap maximum $|\Delta_{max}|$=6.5±0.1meV. Clean and flat Li-termination surfaces (Li-Li unit cell $a_0$=0.38nm) allowed our atomic resolution/register $g(\mathbf{r},\omega)$ measurements to be carried out over the energy range $|\omega|$<30meV (SI Section VI). We then derive the $g(\mathbf{q},\omega)$ in Fig. 2e-h from the measured $g(\mathbf{r},\omega)$ at each energy as shown in Fig. 2i-l. In Fig. 2e we see the expected QPI signature of the anisotropic energy gaps on multiple bands (compare Fig. 2a). Figure 2f shows the characteristic signature of the complete Fermi surface of the γ band of LiFeAs at $\omega$ just outside the superconducting gap edge on that band (compare Fig. 2b). If none of the electron-boson self-energy phenomena intervened, one would expect this closed contour (Fig. 2f) to evolve continuously to smaller and smaller *q*-radius with increasing $\omega$ until the top of this hole-like band is reached. Instead, Fig. 2g shows the beginning of a very different evolution. Above $\omega$~12meV, the **q**-space features become strongly anisotropic in a fashion highly unexpected for un-renormalized states. Indeed, the strongly suppressed $g(\mathbf{q},\omega)$ intensity in the Fe-Fe direction relative to the Fe-As direction is very similar to the predictions for $\Sigma(\mathbf{k},\omega)$ due to AFSF (Fe-Fe direction Fig. 2d).

**8** We compare these results to the predicted $g(\mathbf{q},\omega)$ signatures of a self-energy $\Sigma(\mathbf{k},\omega)$ due to phonons whose strong coupling to electrons is a central premise for the orbital fluctuation scenario (SI Section III). Clearly, comparison of predictions due to the two different boson couplings presented in Fig. 3 through the $\omega$ and $|\mathbf{q}|$ dependence of $g(\mathbf{q},\omega)$ for the Fe-$E_g$ phonon (Fig. 3a-c) and AFSF (Fig. 3d-f) can provide a distinguishing "fingerprint" of AFSF driven effects. The AFSF cause maximum renormalization (peaks of blue curve) in relatively narrow 'beams' in the Fe-Fe directions, precisely where the resonant spin fluctuations are concentrated due to interband scattering (see Fig. 3g). By contrast the electron-$E_g$-phonon interaction is predicted to yield isotropic self-energy signatures (red curve) in QPI data.

**9** In Fig. 4a we show a complete representation of our measured data using a combined **q**-$\omega$ presentation of $g(\mathbf{q},\omega)$ for 0<$\omega$<30meV (ΓX and ΓM **k**-space directions are shown in **q**-space); these data are most clearly demonstrated in SI Movie M1. (Data above $T_c$ and for 0<$\omega$<30meV are shown in SI Section VII.) Most striking in the $g(\mathbf{q},\omega)$ are the anisotropic 'kinks' in $\mathbf{q}(\omega)$ indicated by red arrows. Figure 4b shows the simultaneously measured normalized conductance (~density of states $N(\omega)$), with the characteristic features of pairing interactions indicated by red arrows; these occur within the energy range of the 'kinks' in $\mathbf{q}(\omega)$. Figures 4c-e show plots of $g(\mathbf{q},\omega)$ data along different directions. Figure 4f shows the measured dispersion of the maxima of these $g(\mathbf{q},\omega)$ (SI Section VIII). The inflection points of the $g(\mathbf{q},\omega)$ dispersion seen in Fig. 4a,f, which are directly related to the band renormalization from Re$\Sigma(\mathbf{k},\omega)$, are obviously strongly anisotropic in **q**-space and strongest in the Fe-Fe direction. Finally, Fig. 4g shows measured values of $\Delta E$, the departure of the dispersion of the maxima in $g(\mathbf{q},\omega)$ from a model with no self-energy effect, versus the angle $\theta$ around the $\gamma$-band. This is to be compared with the theoretical prediction in Fig. 3g. The good correspondences between our theoretical prediction for Re$\Sigma(\mathbf{k},\omega)$ effects from coupling to AFSF (Fig. 3g) and the QPI measurements (Fig. 2e-h) are evident. If the optical phonon conjectured to exist in the same energy range is strongly coupling to electrons, a far more isotropic dependence would be expected.

**10** While evidence that self-energy effects due to electron-boson-coupling phenomena are occurring in iron-based materials abounds[26-32], a direct comparison between a theoretical prediction with realistic band/gap structure that distinguishes effects of coupling to AFSF from those due to coupling to $E_g$-phonons generating the orbital fluctuations, has not been achieved. Here, by combining new theoretical insight into QPI discrimination between $\Sigma(\mathbf{k},\omega)$ from resonant-AFSF and $\Sigma(\mathbf{k},\omega)$ due to alternative scenarios, together with novel QPI techniques designed to visualize the $\Sigma(\mathbf{k},\omega)$ signatures[19], we demonstrate that scattering interference at $\omega>\Delta_{max}$ on the $\gamma$ band of LiFeAs is highly consistent with expected effects due to AFSF driven $\Sigma(\mathbf{k},\omega)$. Crucially the apparent changes in the dispersion (Fig. 2 and 4) show strong directional dependence being focused along the Fe-Fe direction where the spin fluctuations of LiFeAs are concentrated[23,33]. This is in excellent qualitative agreement with our predictions based on measured band/gap structures of LiFeAs for resonant AFSF-driven $\Sigma(\mathbf{k},\omega)$ effects (Fig. 2a-d, Fig. 3, SI Section IV). Further, we demonstrate that such anisotropic $\Sigma(\mathbf{k},\omega)$ effects studied here cannot be caused by an Fe-$E_g$ phonon (Fig. 3a-c, SI Section IV). Thus, our combined theory/experiment approach to "fingerprinting" the electronic self-energy $\Sigma(\mathbf{k},\omega)$ discriminates directly between different types of bosonic fluctuations hypothesized to mediate pairing. In analogy to phonon based superconductors, this novel approach may lead to a definite identification of the Cooper pairing mechanism of iron-based superconductivity – with the present result pointing strongly to antiferromagnetic spin fluctuations.

**Acknowledgements**: We are especially grateful to A.P. Mackenzie and D.J. Scalapino for key guidance with this project. We acknowledge and thank D.H. Lee, A. Chubukov, P.J. Hirschfeld, M. Norman, J. Schmalian for helpful discussions and communications. Theoretical studies are supported by the U.S. Department of Energy, Office of Basic Energy Sciences, Division of Materials Science and Engineering under Award DE-SC0010313 (K.L. and E.-A.K.); NSF DMR-0520404 to the Cornell Center for Materials Research (M.H.F.). Experimental studies are supported by the Center for Emergent Superconductivity, an Energy Frontier Research Center, headquartered at Brookhaven National Laboratory and funded by the U.S. Department of Energy, under DE-2009-BNL-PM015; by the UK EPSRC; by a Grant-in-Aid for Scientific Research C (No. 22540380) from the Japan Society for the Promotion of Science. T-M.C. acknowledges support by NSC101-2112-M-001-029-MY3.

**Author Contributions:** M.P.A., A.W.R., F.M. and T.-M.C. performed the experiments and analyzed the data; K.K, A.I, C.-H.L. and H.E synthesized the samples; K.L. and M.H.F. performed the theoretical calculations of the self-energy and simulation of quasiparticle interference. This project was initiated by the experimental discovery of the strongly anisotropic QPI features in the electron-boson energy range (A.W.R.) and by the resulting hypothesis that they are self-energy effects; J.C.D. and E.-A.K. supervised the investigation and wrote the paper with contributions from M.P.A., A.W.R., F.M., K.L. and M.H.F. The manuscript reflects the contributions of all authors.

**Figure Legends**

**Figure 1. Electronic self-energy due to coupling to bosonic fluctuation**

(a) Electronic structure of the first Brillouin zone of FeAs superconductors; here shown using parameters specific to LiFeAs (the inner hole pockets are omitted for clarity). The γ band surrounds the Γ point, the $β_1$ and $β_2$ bands are hybridized surrounding the M point at the corner. The AFSF with $\mathbf{Q}=(π,π)/a_0$ (red arrow) can connect the hole-like bands surrounding the Γ point with the electron-like bands surrounding the M point.

(b) Diagram of the lowest order self-energy contribution from electron-boson interactions.

(c) Spectral function $A(\mathbf{k},ω) \propto \mathrm{Im}G(\mathbf{k},ω)$ of a superconducting hole-like band (with unrenormalized normal-state dispersion as red dashed line) with superconducting gap Δ and the dispersion renormalization at energy Δ+Ω (arrow) due to coupling to a phonon with frequency Ω.

(d) Density of electronic states spectrum $N(ω)$ associated with (c), showing a kink at energy Δ+Ω.

(e) Schematic view of the kinematic constraint in $(\mathbf{k},ω)$-space. We find that the self-energy features on the γ band can only appear at $(\mathbf{k}, ω_\mathbf{k}^γ)$ if there exists a partner point $(\mathbf{k}-\mathbf{Q}, ω_{\mathbf{k}-\mathbf{Q}}^m)$ with $ω_{\mathbf{k}-\mathbf{Q}}^m = ω_\mathbf{k}^γ - Ω \geq Δ$ to satisfy the kinematic constraint. The blue surface at the center and the yellow surfaces at the corners of the Brillouin zone are defined by the hole-band and the outer electron band dispersion. The red surface indicates the hole-band displaced by the AFSF momentum $\mathbf{Q}=(π,π)/a_0$ (dark red arrow) and energy Ω (light red arrow). The points that satisfy the kinematic constraint (equation (2)) are defined by the intersection of the red and blue surfaces, and indicated with a solid black line. These points are expected to exhibit the strongest self-energy effects due to coupling to AFSF. The anisotropy of the black line demonstrates directly how the AFSF self-energy effects must exist at different ω in different $\mathbf{k}$-space directions around a particular Fermi pocket (e.g. γ band in Fig. 1a).

**Figure 2. Comparison between scattering interference theory with AFSF driven self-energy effects and the experiments.**

(a-d) Theoretically predicted QPI patterns $g(\mathbf{q},\omega)$ for LiFeAs with Green's function including the self-energy effect due to the coupling between electrons and resonant AFSF fluctuations as described in SI Section II. In these simulations, we suppressed the interband scattering visible in the data to highlight the QPI of the γ band that are the focus of this study. Note in (c,d) the strong anisotropy induced by the kinematic constraint (equation (2)) with clear suppression of $g(\mathbf{q},\omega)$ for $\mathbf{q}$ along Fe-Fe direction, which is strikingly different from the strong gap anisotropy that dictates the pattern in (a).

(e-h) Measured QPI patterns $g(\mathbf{q},\omega)$ (obtained from $g(\mathbf{r},\omega)$ of LiFeAs). (e), QPI signature of anisotropic energy gaps; (f), Expected isotropic signature of the complete Fermi surface of the γ band; (g,h) show the transition to a strongly anisotropic $g(\mathbf{q},\omega)$. Note the suppression of $g(\mathbf{q},\omega)$ occurring along Fe-Fe direction.

(i-l) Real space images of $g(\mathbf{r},\omega)$ from which (e-h) were obtained. The insets show a zoom-in onto a particular impurity revealing the real space standing waves from QPI

**Figure 3. "Fingerprint" distinguishing antiferromagnetic spin-fluctuations from phonon generated orbital-fluctuations in LiFeAs**

(a-c) Predicted QPI response calculated with self-energy driven by the Fe-$E_g$ phonon. (a,b), Sequential images of $g(\mathbf{q},\omega)$ for two different $\omega$, one below and one near the coupling energy. (c), Predicted $g(\mathbf{q},\omega)$ in three different directions in $\mathbf{q}$-space, corresponding to the Fe-As direction (left), Fe-Fe direction (right), and in between (center). Different grey lines correspond to different $\omega$, with 1meV increase between each neighboring pair, starting from the lowest bias $\omega=0$ at the bottom. The plots are offset for clarity, and the red dots indicate the maxima. The $g(\mathbf{q},\omega)$ on the γ band remains virtually isotropic, despite the momentum dependence of the electron-phonon coupling in our simulations.

(d-f) Predicted QPI response calculated with self-energy driven by resonant AFSF.

(d,e), predicted $g(\mathbf{q},\omega)$ for the same energies as in (a,b). (f) Predicted $g(\mathbf{q},\omega)$ in three different directions in **q**-space as in (c). $g(\mathbf{q},\omega)$ on the γ band is predicted to be highly anisotropic.

(g)   Predicted $\mathrm{Re}\Sigma(\mathbf{k}(\omega,\theta),\omega)$ at fixed energy $\omega$=10meV calculated with self-energy driven by resonant AFSF (blue) and the Fe-$E_g$ phonon (red) as a function of the angle $\theta$ (as defined in (b,e)) around the γ band.

**Figure 4. QPI measurements of anisotropic renormalization of dispersion due to self-energy in LiFeAs**

(a)   Measured $g(\mathbf{q},\omega)$ represented in $\mathbf{q}$-$\omega$ space for 0<$\omega$<30meV, with the (0,1) and (1,1) directions highlighted. The inset shows the measured data up to $\omega$=60meV. Red arrow indicates the energy $\omega$~12meV at which sudden changes in dispersion and isotropy of $g(\mathbf{q},\omega)$ are observed. See SI Movie 1 in which this effect is vivid.

(b)   The $N(\omega)$ measured simultaneously with $g(\mathbf{q},\omega)$ and normalized by $N(\omega)$ at $T$=16K. Vertical red arrows indicate the energy $\omega$~12meV where features associated with Cooper pairing are observed. The inset shows the original $N(\omega)$~$dI/dV(\omega)$

(c-e)  Lineplots of measured $g(\mathbf{q},\omega)$ data for different energies $\omega$ along the Fe-As direction (left), the Fe-Fe direction (right), and in between. The data at different $\omega$ are offset vertically for clarity. The angle indicated is $\theta$ measured from the Fe-As direction. The red lines represent fits as discussed in SI Section VIII.

(f)   Dispersion of the maxima in $g(\mathbf{q},\omega)$ extracted from line cuts as in (c-e) (SI Section VIII). The angle indicated is $\theta$ measured from the Fe-As direction. These dispersions have to be compared with the predictions in Fig. 3a-c or Fig. 3d-f.

(g)   Measured $\Delta E$, the departure of the dispersion of the maxima in $g(\mathbf{q},\omega)$ from a model with no self-energy effect, as a function of the angle $\theta$ around the γ band of LiFeAs. This is to be compared with the theoretical prediction in Fig. 3g.

Figure 1

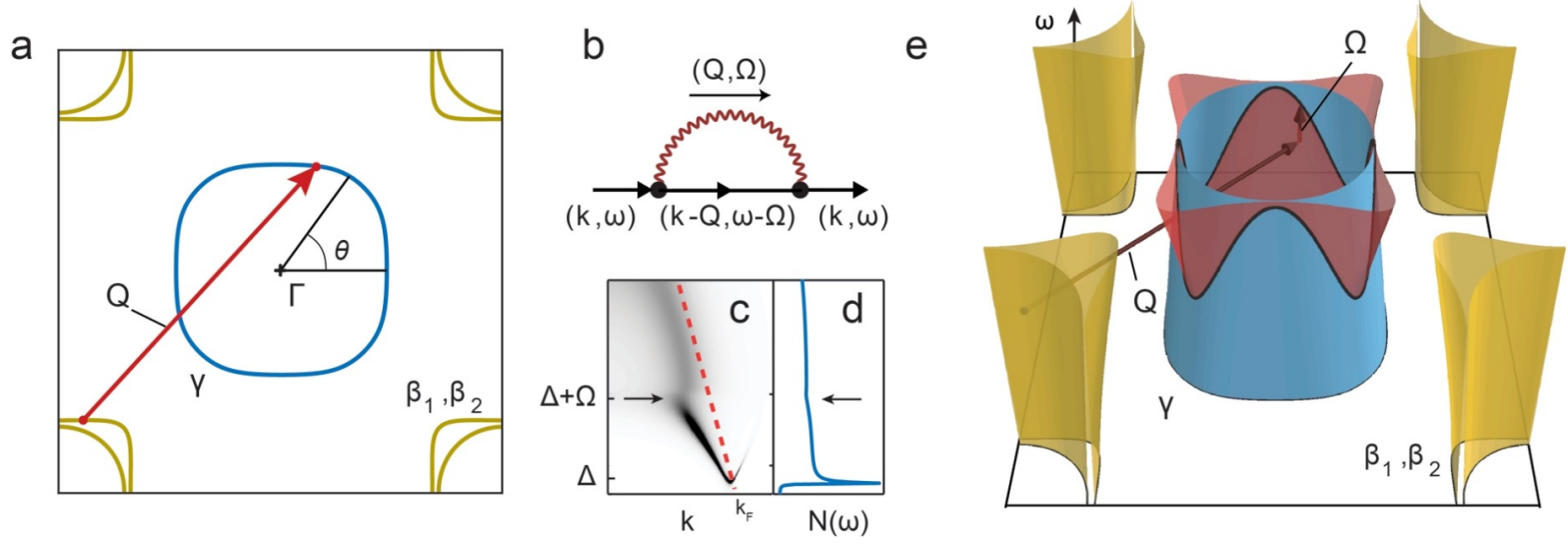

*Figure 2*

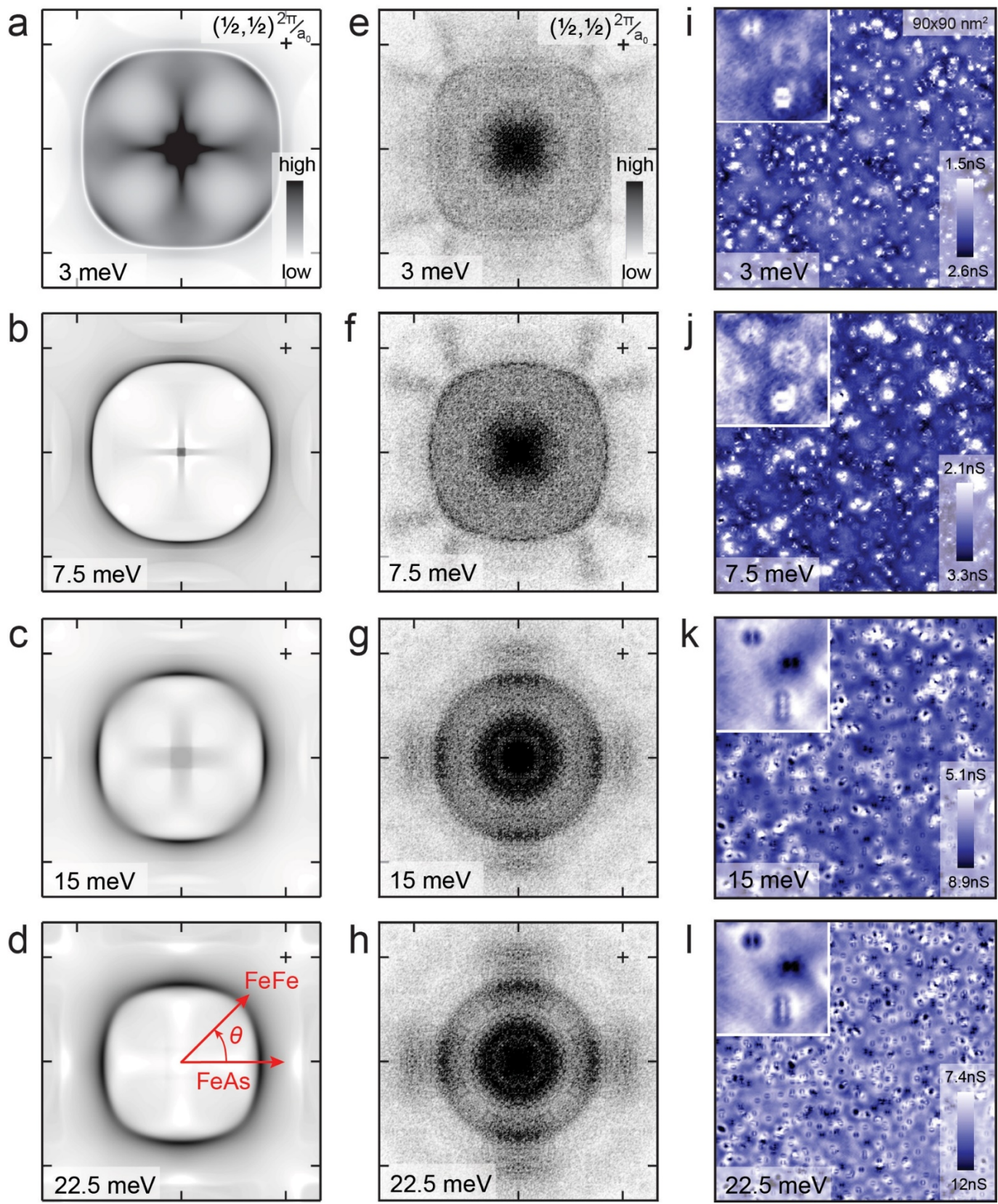

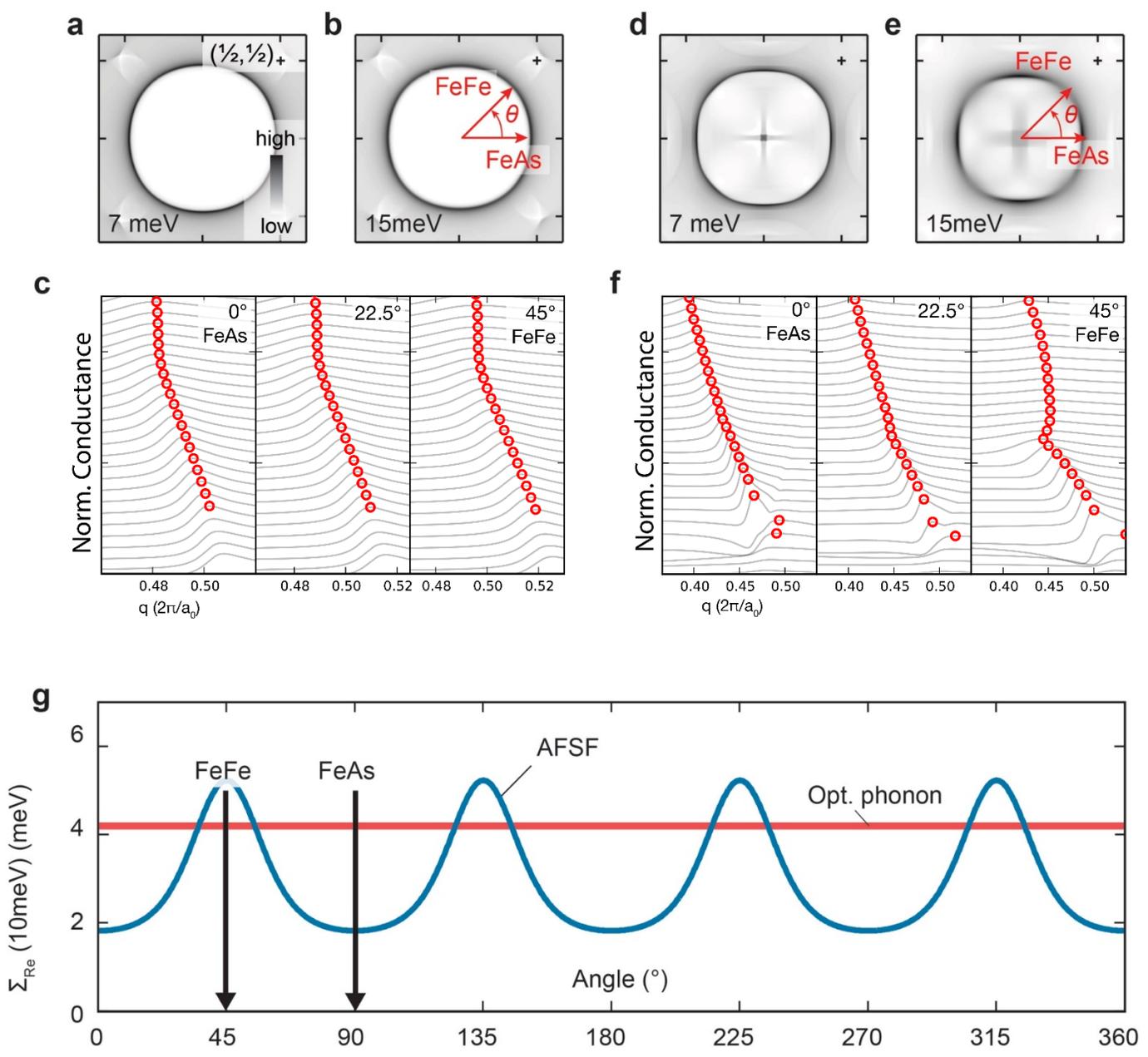

Figure 4

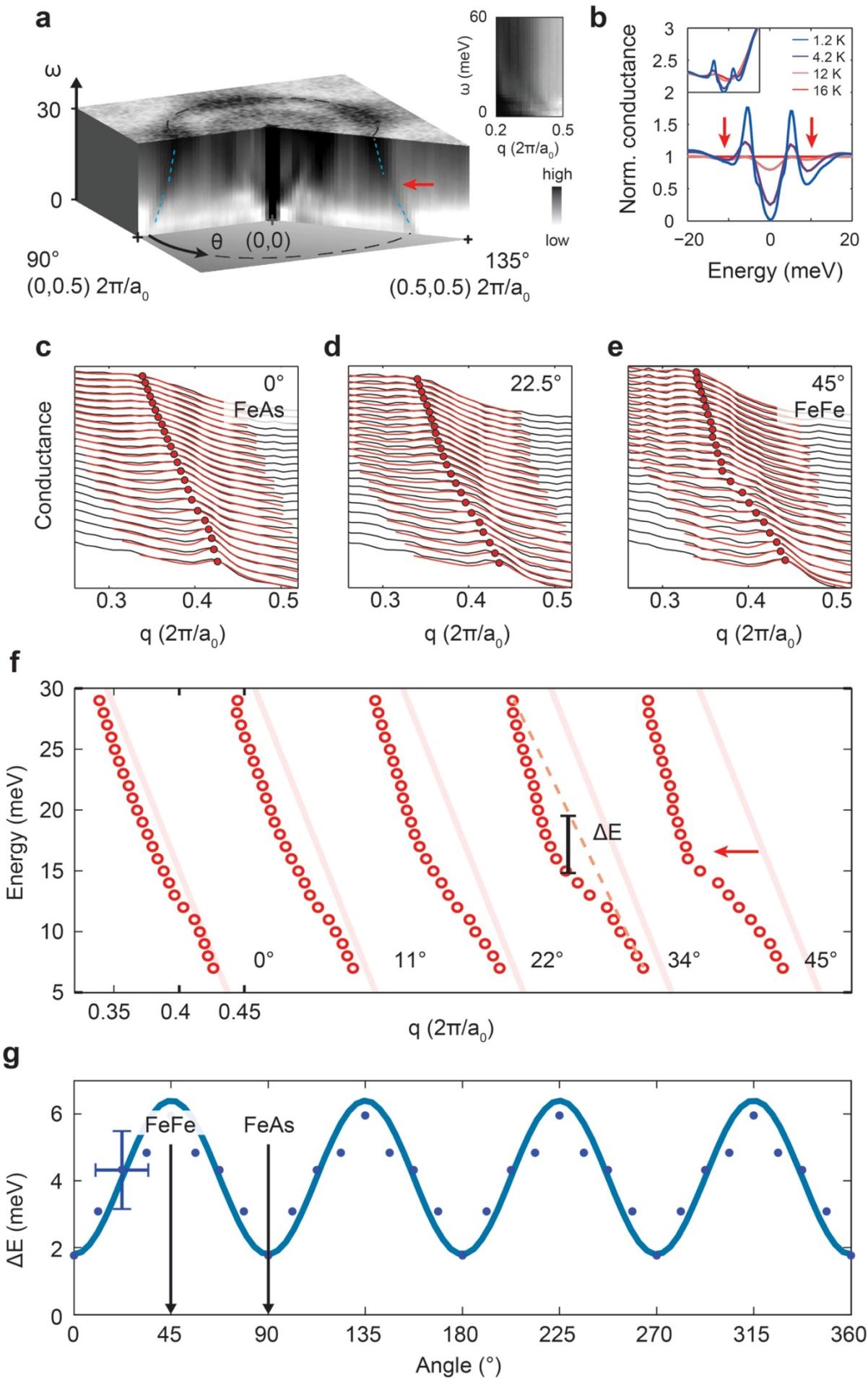